\documentclass[aps, prd, twocolumn, floatfix, superscriptaddress, longbibliography, groupedaddress]{revtex4-2}
\pdfoutput=1

\usepackage{amsmath, amssymb}

\usepackage{graphicx}
\usepackage{color}
\usepackage{hyperref}

\usepackage{booktabs}

\usepackage{siunitx}

\usepackage{scalerel} 

\AtBeginDocument{
  \heavyrulewidth=.08em
  \lightrulewidth=.06em
  \cmidrulewidth=.03em
  \belowrulesep=.65ex
  \belowbottomsep=0pt
  \aboverulesep=.4ex
  \abovetopsep=0pt
  \cmidrulesep=\doublerulesep
  \cmidrulekern=.5em
  \defaultaddspace=.5em
  }

\newcommand{\papertitle}{Space-time symmetry and nonreciprocal parametric resonance in mechanical systems}
\newcommand{\acktext}{We thank Yogesh Joglekar for useful discussions. This work was supported by the National Science Foundation under Grant No. CMMI-2128671 and Grant No. DMR–2145766.}

\newcommand{\eqnref}[1]{Eq.~\eqref{#1}}
\newcommand{\Equationref}[1]{Equation~\eqref{#1}}
\newcommand{\appref}[1]{Appendix~\ref{#1}}
\newcommand{\subfiglabel}[1]{{\bfseries #1}}
\newcommand{\figref}[1]{Fig.~\ref{#1}}
\newcommand{\subfigref}[2]{\figref{#1}\subfiglabel{#2}}
\newcommand{\secref}[1]{Sec.~\ref{#1}}
\newcommand{\bs}[1]{\boldsymbol{#1}}

\newcommand\id{\mathbb{I}_n}
\newcommand\bigz{0}
\newcommand{\ket}[1]{{| #1 \rangle}}
\newcommand{\bra}[1]{{\langle #1 |}}
\newcommand{\expect}[1]{{\langle #1 \rangle}}

\newcommand{\spt}{X_n}
\newcommand{\inte}{r}

\begin{document}
\title{\papertitle}
\author{Abhijeet Melkani}
\email{amelkani@uoregon.edu}
\affiliation{Institute for Fundamental Science, Materials Science Institute, and Department of Physics,
University of Oregon, Eugene, OR 97403}

\author{Jayson Paulose}
\email{jpaulose@uoregon.edu}
\affiliation{Institute for Fundamental Science, Materials Science Institute, and Department of Physics,
University of Oregon, Eugene, OR 97403}

\begin{abstract}
Linear mechanical systems with time-modulated parameters can harbor oscillations with amplitudes that grow or decay exponentially with time due to the phenomenon of parametric resonance.
While the resonance properties of individual oscillators are well understood, those of systems of coupled oscillators remain challenging to characterize.
Here, we determine the parametric resonance conditions for time-modulated mechanical systems by exploiting the internal symmetries arising from the real-valued and symplectic nature of classical mechanics.
We also determine how these conditions are further constrained when the system exhibits external symmetries. 
In particular, we analyze systems with space-time symmetry where the system remains invariant after a combination of discrete translation in both space and time.
For such systems, we identify a combined space-time translation operator that provides more information about the dynamics of the system than the Floquet operator does, and use it to derive conditions for one-way amplification of traveling waves.
Our exact theoretical framework based on symmetries enables the design of exotic responses such as nonreciprocal transport and one-way amplification in dynamic mechanical metamaterials, and is generalizable to all physical systems that obey space-time symmetry.
\end{abstract}

\maketitle

\section{Introduction}

Parametric resonance---the injection of energy into an oscillator through periodic time-modulation of a system parameter---is a fundamental physical mechanism that compromises the stability of mechanical structures~\cite{landau1982mechanics,YakubovichStarzhinskii} but can also be exploited to amplify electrical~\cite{louisell1960coupled, huertamorales2023synthetic} and mechanical~\cite{rugar1991squeezing, mathew2016parametric,trainiti2019filtering} signals.
When combined with nonlinear oscillator dynamics, parametric resonance underpins the working principles of classical time crystals~\cite{heugel2019timeCrystal, yao2020timeCrystal}, coherent Ising machines~\cite{wang2013parametricIsing, inagaki2016ising, bello2019coherentBeating}, quantum microwave amplifiers~\cite{roy2016parametric}, and other mechanical and NEMS-based devices~\cite{buks2002micromechanical, matheny2019exotic, kim2023nonlinear}.
Engineering parametric resonances by patterning system parameters in space and time pave a route towards active metamaterials with tunable signal processing capabilities~\cite{zangeneh-nejad2019review,wang2020tunable}.

For a single oscillator, the existence of parametric resonances is governed by Hill's equation~\cite{magnus1966hill} (the Mathieu equation being a special case~\cite{ruby1996mathieu, kovacic2018mathieu}), which predicts resonances at modulation time-periods that are near integer multiples of half the oscillator's natural time period~\cite{broer1995geometrical, YakubovichStarzhinskii}.
In contrast, the resonance conditions for arbitrary systems of coupled parametric oscillators are sensitive to the modulation phases on individual oscillators~\cite{salerno2016modulation, calvanese2019coupled, kruss2022oneway}. 
Resonances are typically revealed only after a numerical or perturbative calculation of the Floquet matrix (the time-evolution operator over one modulation period).

Here, we derive the resonance conditions of coupled parametric oscillators by harnessing recent advances in the symmetry analysis~\cite{ashida2020nonhermitian} and topological classification~\cite{liu2019topodefects, kawabata2019symmetry, zhou2019topology} of non-Hermitian quantum systems.
In doing so, we extend previous work on the topological classification of static mechanical systems via classical-quantum mapping~\cite{susstrunk2016phonons, kane2014topological, ashida2020nonhermitian} to time-dependent systems by using non-Hermitian and Floquet techniques.
Specifically, we establish parametric resonance as an example of a real-to-complex eigenvalue transition or pseudo-Hermiticity breaking~\cite{mostafazadeh2002pseudoHermitian, melkani2023breaking} in the Floquet matrix.
Our resonance conditions are formulated in terms of the symmetries obeyed by the time-modulated system and thus do not rely on specific functional forms of the parametric modulation.
Furthermore, we demonstrate that the static limit of the Floquet matrix, which can be evaluated exactly from the unmodulated system, is sufficient to reveal these conditions.

Upon including external symmetries, the space of potential outcomes for parametric resonance is significantly enriched.
The effect of static external symmetries (such as discrete translation symmetry) is straightforward---vector spaces associated with different symmetry eigenvalues decouple, thereby further constraining the spectra and their topological classification~\cite{kane2014topological,susstrunk2016phonons}.
However, time-modulated systems can admit space-time symmetries~\cite{xu2018spaceTimeGroup} in which the system remains invariant after combined discrete translations in space and time.
For such a symmetry, the time-dependent Hamiltonian no longer splits into decoupled blocks.
Instead, we identify a general framework in terms of an operator combining spatial translation with time evolution, that captures more information about the system dynamics than the Floquet matrix does, to analyze the dynamics of space-time symmetric systems. 

\begin{figure}[tb]
\includegraphics{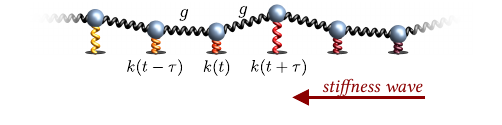}
\caption[]{\label{fig:coupledSystem} 
A system of $n$ oscillators (shown with random displacements) of unit mass connected to their nearest neighbours via springs of stiffness $g$ (black springs) and to the ground with a spring of stiffness $k_j(t) = k[1+\delta f(t + j\tau)]$ (colored springs). 
Here, $j$ indexes the oscillators, $k$ is the grounding stiffness in the absence of time-modulation, $\delta$ sets the modulation strength, $f(t)$ is a $T$-periodic function, and $\tau$ is the modulation offset between nearest neighbours. 
We assume periodic boundary conditions in the system. 
When $\tau=\frac{T}{n}$, the system enjoys space-time symmetry, i.e., it is invariant after a translation by one position in the leftwards direction followed by a time evolution of $\tau$ time units.
Equivalently, a stiffness wave travels in the direction of decreasing $j$ (from dark red to yellow).
}
\end{figure}

We demonstrate the utility of this operator, which we term the \emph{space-time Floquet matrix}, by applying it to a ring of coupled parametric oscillators satisfying space-time symmetry (\figref{fig:coupledSystem}).
We show that several normal modes of the unmodulated ring are protected by symmetry from resonating at frequencies where parametric modulation would naively be expected to occur.
For example, the lowest-frequency or ground-state mode of the unmodulated system has frequency $\omega$, the natural frequency of the individual oscillators.
Yet, the smallest time-period of modulation at which the ground-state normal mode becomes resonant is $T=n\pi/\omega$ (rather than the resonance period $T=\pi/\omega$ of each oscillator), where $n$ is the number of oscillators in the ring.
For a large system, the ground state is effectively protected from parametric resonance in a space-time modulated system.

The space-time Floquet matrix also determines the conditions for amplification of one-way propagating modes. 
In the ring of oscillators, wavelike modes occur in degenerate pairs which travel in opposite directions.
We show that the traveling-wave parameter modulation amplifies these counterpropagating modes at different frequencies, thereby breaking left-right symmetry.
The conditions for this selective one-way amplification are obtained exactly from the symmetries alone, independently of the functional form of the time-modulation. 
These conditions provide a simple way to engineer nonreciprocal transport~\cite{nassar2020nonreciprocity} and one-way amplification~\cite{galiffi2019broadband, kruss2022oneway} in non-Hermitian Floquet systems.

Our results provide a theoretical framework, based on non-Hermitian and Floquet symmetries, to design collective mechanical states with interesting resonance and transport properties. Potential applications include sensing~\cite{blaikie2019bolometer, budich2020sensors,wiersig2022response}, active signal processing~\cite{zangeneh-nejad2019review, wang2020tunable, trainiti2019filtering, kim2023nonlinear, matheny2019exotic, bender2013PToscillators}, and computing~\cite{wang2013parametricIsing, inagaki2016ising}. 
In addition, the space-time Floquet matrix provides an exact framework for analyzing all systems with space-time symmetry~\cite{galiffi2019broadband, oudich2019spaceTime, gao2021spaceTimeTunneling}, beyond the mechanical systems that are the focus of this work. 

This paper is structured as follows. 
In \secref{sec:theory} we analyze the internal symmetries of linear classical mechanical systems. 
We begin by developing a framework for static systems in \secref{subsec:time-independent} which we then generalize to time-dependent systems in \secref{subsec:Floquet}, identifying the conditions for parametric resonance. The effects of external symmetries are analyzed in \secref{sec:sym}, specifically translational symmetry (\secref{subsec:sym1}) and space-time symmetry (\secref{subsec:sym2} and \secref{subsec:sym3}). 
We discuss the significance of our results and possible future directions in \secref{sec:discussion}.

\section{Non-Hermitian Floquet theory of parametric oscillators}\label{sec:theory}

We describe the stability and mode structure of coupled parametric oscillators aided by the  symmetries and topological structures that have been identified in non-Hermitian and Floquet quantum systems~\cite{liu2019topodefects,kawabata2019symmetry, zhou2019topology}. 
Since these techniques apply to systems that are first-order in time, we use a Hamiltonian formulation of the dynamics of coupled oscillators, which we first describe for oscillators with static parameters before generalizing to the time-dependent case. 
Different variants of the time-independent formulation below have appeared in Refs.~\cite{susstrunk2016phonons, yoshida2019ring, melkani2023breaking}.

\subsection{Time-independent case} \label{subsec:time-independent}

Consider $n$ coupled classical mechanical oscillators of equal masses (set to $1$). 
Denote the positions of the oscillators by $\bs x = (x_1, x_2, \dots, x_n)^T$ such that the potential energy of the system is $\bs{x}^T\cdot K \cdot\bs x$ where $K$ is the stiffness matrix (a.k.a. the dynamical matrix). 
The stiffness matrix is real and symmetric, with real eigenvalues $\Omega_i^2$ and corresponding eigenvectors (normal modes) $\bs q_i$ satisfying
\begin{equation}
    K \bs q_i = \Omega_i^2 \bs q_i.
\end{equation}
In the absence of dissipation or velocity-dependent forces, the equation of motion of the oscillators is
\begin{equation} \label{eq:eom}
i\frac{d}{dt} \begin{pmatrix} \bs x(t)\\ \bs p(t) \end{pmatrix} 
= -i\begin{pmatrix} 
\bigz & -\id \\ 
K &  \bigz 
\end{pmatrix}
\begin{pmatrix} \bs x(t) \\ \bs p(t) \end{pmatrix}
\end{equation}
where $\bs p = (p_1, p_2, \dots, p_n)^T$ denotes the momenta of the oscillators.
While we disregard friction and velocity-dependent forces such as the Lorentz force or gyroscopic forces here, in \appref{app:generalizations} we show that the results below generalize as long as the friction is uniform for all oscillators and the velocity-dependent  forces have no explicit time dependence. 

\Equationref{eq:eom}  defines the quantum Hamiltonian
\begin{equation}\label{eq:hamiltonianDefinition}
    H = -i\begin{pmatrix} \bigz & -\id \\ K & \bigz \end{pmatrix}
\end{equation}
in terms of the classical stiffness matrix.
When the Hamiltonian is time-independent the equation can be solved by substituting $\begin{pmatrix} \bs x (t)\\ \bs p(t) \end{pmatrix} = e^{-i\omega t} \ket{v}$, where $\ket v$ is a time-independent column-vector, to get
\begin{equation}
\omega \ket{v} = H \ket{v},
\end{equation}
an eigenvalue equation. The eigenvectors are
\begin{equation}\label{eq:oscillatorvectors}
\ket{v_i^{\pm}} =\begin{pmatrix} \bs q_i \\ -i\omega_{i}^{\pm} \bs q_i \end{pmatrix}
\end{equation}
with eigenvalues
\begin{equation}\label{eq:oscillatorvalues}
    \omega_i^{\pm} = \pm \Omega_i.
\end{equation}
The partners $\ket{v_i^+}$ and $\ket{v_i^-}$ are associated with the same normal mode $\bs{ q_i}$ but differ in the phase relationship between position and velocity.

The Hamiltonian $H$, generically, has two internal symmetries. 
First, since classical mechanics is real-valued, by our choice of notation we have $H^* = -H$ such that its eigenvalues are either purely imaginary or come in pairs with oppositely signed real parts. 
Second, $H$ is pseudo-Hermitian, $ G H G^{-1} = H^\dagger$, where
\begin{equation}\label{eq:symplectic}
    G = iJ = i\begin{pmatrix} \bigz & \id \\ -\id & \bigz \end{pmatrix}
  \end{equation}
is the intertwining operator which is Hermitian as well as unitary. 
This pseudo-Hermiticity (a generalization of $\mathcal{PT}$ symmetry~\cite{bender2019PTbook}) arises from the symplectic structure of Hamilton's equations of motion~\cite{melkani2023breaking} ($J$ is the symplectic form). Pseudo-Hermiticity implies that the eigenvalues of $H$ are either real or come in complex-conjugate pairs. 

Taking the two symmetries together, $H$ must either have pairs of oppositely signed eigenvalues which are purely real or purely imaginary, or it must have quadruplets of eigenvalues with non-zero real as well as imaginary parts forming sets of the form, $\{\lambda, -\lambda, \lambda^*, -\lambda^*\}$. We will show that these facts hold true even in the time-dependent case.

For the energy of the system to be non-negative, the stiffness matrix $K$ is constrained to be positive-definite (in the absence of zero-frequency modes, also called zero modes). 
While this constraint does not introduce any additional non-Hermitian internal symmetries, it enforces the eigenvalues $\omega_i^{\pm} = \pm \Omega_i$ to be purely real. 
The Hamiltonian $H$ can now be transformed, via a non-unitary similarity transformation, to a Hermitian matrix enabling the use of the Hermitian topological classification for conservative mechanical systems~\cite{susstrunk2016phonons}. 
This transformation requires taking the square-root of the matrix $K$; possible complications due to branch-points are avoided if zero modes are excluded. 
In the language of the Hermitian topological classification, the two symmetries noted above are mapped to the Hermitian time-reversal symmetry and the Hermitian chiral symmetry respectively~\cite{susstrunk2016phonons}. 

In the presence of zero modes, an alternate strategy is to use the equilibrium matrix, which links displacements to strains, in place of the dynamical matrix~\cite{kane2014topological,mao2018maxwell}. 
In either case, the topological classification in terms of Hermitian matrices is not applicable to either time-dependent systems or dissipative systems where the frequency eigenvalues are not constrained to be purely real. 
Non-Hermitian symmetries, as used in this work, are arguably more useful since they generalize easily, as we shall see below, to such non-conservative systems~\cite{yoshida2019ring, ghatak2020mechanical, wang2023staticnonHermitian}. In the GBL classification scheme of non-Hermitian Hamiltonians~\cite{liu2019topodefects}, the first internal symmetry we identified corresponds to the $K$ symmetry: $H=\epsilon_k k H^* k^{-1}, kk^* = \eta_k \mathbb{I}$ with $\epsilon_k = -1$ and $\eta_k = +1$. The second internal symmetry corresponds to the $Q$ symmetry: $H=\epsilon_q q H^\dagger q^{-1}, q^2 = \mathbb{I}$ with $\epsilon_q = +1$.

A real eigenvalue of a pseudo-Hermitian matrix $H$ with intertwining operator $G$ can be classified by its Krein signature, which is given by the sign of $\expect{v|G|v}$~\cite{melkani2023breaking, YakubovichStarzhinskii, coppel1965stability}. 
Here $\ket{v}$ is the corresponding eigenvector of $H$ and $\bra{v} = \ket{v}^\dagger$ the conjugate-transpose of the eigenvector.  
Upon tuning some parameter in the matrix, these real eigenvalues can collide on the real axis and turn into complex-conjugate eigenvalues (a phenomenon known as pseudo-Hermiticity breaking) only if they have opposite signatures~\cite{melkani2023breaking}.
The collision typically marks a singularity in the eigenvalue spectrum called an exceptional point, at which the matrix becomes non-diagonalizable~\cite{bender2019PTbook}.
Pseudo-Hermiticity breaking is typically accompanied by drastic changes in the behavior of a physical system.
Examples include the emergence of amplified/damped modes in $\mathcal{PT}$-symmetric systems~\cite{elganainy2018nature} and unusual phase transitions driven by exceptional points~\cite{hatano1996localizationprl, hamazaki2019localization, fruchart2021nonreciprocal, melkani2021polymers}.
In the mechanical system under consideration, pseudo-Hermiticity breaking generates one or more frequencies with positive imaginary component, corresponding to amplified or unstable modes.  

To find the Krein signature of each eigenvalue of our Hamiltonian, we compute
\begin{equation}\label{eq:kindComputation}
\langle v_i^{\pm}| G| v_i^{\pm}\rangle = 
\begin{pmatrix} \bs{q_i}^* , i\omega_{i}^{\pm} \bs{q_i}^* \end{pmatrix} G
\begin{pmatrix} \bs q_i \\ -i\omega_{i}^{\pm} \bs q_i \end{pmatrix} = \pm 2|\bs q_i|^2 \Omega_i,
\end{equation}
which shows that the eigenspaces corresponding to $\ket{v_i^{+}}$ have positive signature while those corresponding to $\ket{v_i^{-}}$ have negative signature. 
Physically, these two signatures are distinguished by whether the momenta are lagging behind or ahead of the positions (or, in the case of parametric modulation, whether they are in phase or out of phase with the modulation). 
In the absence of dissipation or parametric modulation, eigenvalues of different signatures can meet only at the origin, i.e. at $\omega = 0$.
Indeed the zero modes (or floppy modes) govern the instability of static mechanical systems~\cite{kane2014topological}. 

The Krein classification of eigenvalues can be performed even for a time-dependent system and will be crucial for the determination of the conditions for parametric resonance.

\subsection{Time-dependent case via Floquet theory}\label{subsec:Floquet}

We now consider the effect of modulating the system parameters---specifically, the spring stiffnesses---so that the stiffness matrix $K(t)$ is time-dependent.
The equations of motion now take the form of the time-dependent Schr\"odinger equation,
\begin{equation}\label{eq:timeDependent}
i\frac{d}{dt} \begin{pmatrix} \bs x(t)\\ \bs p(t) \end{pmatrix} 
= -i\begin{pmatrix} 
\bigz & -\id \\ 
K(t) & \bigz 
\end{pmatrix}
\begin{pmatrix} \bs x(t) \\ \bs p(t) \end{pmatrix} = H(t)\begin{pmatrix} \bs x(t) \\ \bs p(t) \end{pmatrix}.
\end{equation}
The solution to this equation is
\begin{equation}
\begin{pmatrix} \bs x(t) \\ \bs p(t) \end{pmatrix} = U(t)\begin{pmatrix} \bs x(0) \\ \bs p(0) \end{pmatrix},
\end{equation}
where $\begin{pmatrix} \bs x(0) \\ \bs p(0) \end{pmatrix}$ are the initial conditions of the oscillators and the time-propagation matrix $U(t)$ satisfies
\begin{equation} \label{eq:timeprop}
i\frac{d}{dt} U(t) = H(t)U(t)
\end{equation}
with the initial condition $U(0)=\mathbb{I}_{2n}$. 

\begin{figure}[tb]
\includegraphics{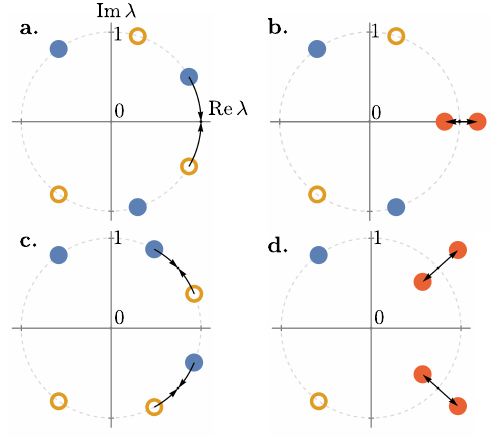}
\caption[]{\label{fig:krein} 
  Pseudo-Hermiticity breaking in the spectrum of the time-propagation matrix $U(t)$. 
  \subfiglabel{a.} Representative spectrum for a system with three degrees of freedom, with all eigenvalues of $U(t)$ lying on the unit circle. These eigenvalues form pairs $\{\lambda = 1/\lambda^*, \lambda^* = 1/\lambda \}$ with positive (solid, blue)  and negative (hollow, yellow) Krein signature.
  Upon tuning a parameter of the system, isolated eigenvalues move along the unit circle (solid arrows).
  \subfiglabel{b.} When a pair of eigenvalues of opposite Krein signature collide, they can move off the unit circle (solid, red), signifying pseudo-Hermiticity breaking. They are constrained to lie on the real axis to satisfy $\{ \lambda = \lambda^*, 1/\lambda = 1/\lambda^* \}$.
  \subfiglabel{c.} Same as \subfiglabel{a,} but with two pairs of eigenvalues colliding away from the real axis.
  \subfiglabel{d.} After pseudo-Hermiticity breaking, the two colliding pairs move off the unit circle as a quartet of four distinct values $\{\lambda, \lambda^*, 1/\lambda, 1/\lambda^* \}$.
}
\end{figure}

The spectrum of the time-propagation matrix exhibits features inherited from the internal symmetries satisfied by $H(t)$.
First, the condition $H^*(t) = -H(t)$ implies that the time-propagation matrix is real,
\begin{equation}
U^*(t) = U(t).
\end{equation}
As a result, its eigenvalues are either purely real or they come in complex-conjugate pairs. Second, the symplecticity condition, $ G H(t) G^{-1} = H^\dagger(t)$, gives~\cite{liu2022floquet}
\begin{equation}
U^{-1}(t) = G^{-1}U^\dagger(t) G.
\end{equation}
This symmetry ensures that eigenvalues of $U(t)$
come in reciprocal pairs (or are either of $+1$ or $-1$).
The two symmetries together require that the eigenvalues of
$U(t)$
appear in sets of the form $\{\lambda, \lambda^*, 1/\lambda, 1/\lambda^* \}$. 

Analogous to the time-independent case, we note that the matrix $M(t):= i \log[U(t)]$ satisfies the pseudo-Hermiticity condition $ G M(t) G^{-1} = M^\dagger(t)$. 
The eigenvectors $\ket{v_i}$ of $M(t)$ are the same as that of $U(t)$ while the corresponding eigenvalue is $i\log(\lambda_i)$ which is real if $\lambda_i$ lies on the unit circle of the complex plane.   
Thus, each eigenvalue $\lambda_i$ of $U(t)$ which lies on the unit circle can be assigned a Krein signature according to the sign of $\expect{v_i|G|v_i}$ where $\ket{v_i}$ is the corresponding eigenvector of $U(t)$. 
Upon smoothly varying any parameter that affects $U(t)$ (such as any spring stiffness in the stiffness matrix or the time variable itself), isolated eigenvalues of $U(t)$ on the unit circle cannot move off it unless two eigenvalues of opposite Krein signature collide~\cite{YakubovichStarzhinskii}. 
The collision signifies pseudo-Hermiticity breaking in $M(t)$ which we equivalently refer to as pseudo-Hermiticity breaking in $U(t)$. 
Two examples of such pseudo-Hermiticity breaking transitions are shown schematically in \subfigref{fig:krein}{a--b} and \subfigref{fig:krein}{c--d}.

We now restrict the parametric modulation to be periodic in time with time period $T$, so that $K(t+T) = K(t)$.
By Floquet's theorem, the behavior of the system at time scales much longer than the modulation period is determined by the eigenvalues (also called \emph{Floquet multipliers}) $\lambda_i(T)$ of the \emph{Floquet matrix} $U(T)$ (see \appref{app:floquet} for details on Floquet methods).
In particular, when a system is initialized with the $i$th eigenvector $\ket{v_i}$ of $U(T)$ (called a \emph{Floquet eigenvector}) at $t=0$, its state at integer multiples of the time period is $\begin{pmatrix} \bs x(mT) \\ \bs p(mT) \end{pmatrix} = U^m(T)\ket{v_i}= \lambda_i^m \ket{v_i}$.
Eigenvalues of $U(T)$ that lie off the unit circle correspond to modes that grow or decay exponentially with time and indicate parametric resonances.

The pseudo-Hermiticity breaking structure sketched in \figref{fig:krein} can be used to identify conditions for the existence of parametric resonances when the strength of the parametric modulation is small, i.e., when $K(t) = K_0 + \delta K_1(t)$ with $K_1(t+T) = K_1(t)$ and $\delta \ll 1$.
In the limiting case, the eigenvalues of $U(t; \delta \rightarrow 0)\sim e^{-iH t}$ are $e^{\mp i \Omega_i t}$ where $\pm \Omega_i$ are precisely the eigenvalues of the time-independent Hamiltonian in \eqnref{eq:hamiltonianDefinition}, with Krein signatures identified in \eqnref{eq:kindComputation}.
As time advances, the eigenvalues of $U(t)$ with positive (negative) Krein signature move clockwise (counter-clockwise) along the unit circle (see \subfigref{fig:spaceTime}{b} for an example involving a system with two degrees of freedom).
Eigenvalues of opposite Krein signature coincide when $e^{ -i\Omega_i t} = e^{+i \Omega_j t}$ for some $i$ and $j$ (the case of $i=j$ included), which occurs when
\begin{equation}\label{eq:generalCondition}
t = \inte \frac{2\pi}{\Omega_i+\Omega_j}
\end{equation}
where $\inte$ (here and throughout the paper) denotes an arbitrary positive integer.
Since such collisions are the precursor to the eigenvalues moving off the unit circle, they are termed unstable degeneracies.

When parametric modulation is turned on ($\delta > 0$) the eigenvalues of $U(t; \delta)$ are approximately equal to those of $U(t;\delta \to 0)$ by continuity~\footnote{Continuity of $U(t;\delta)$ as a function of $\delta$ is a consequence of the continuity of $H(t;\delta)$ as a function of $\delta$ (see Section II.1.3 of Ref.~\cite{YakubovichStarzhinskii})}.
Generically (in the absence of additional symmetries inhibiting them), they move off the unit circle at values of time near the unstable degeneracies, \eqnref{eq:generalCondition}.
If the system is parametrically modulated with such time-periods, these eigenvalues moving off the unit circle would precisely be the Floquet multipliers.
Thus parametric resonance is expected whenever the modulation time-period $T$ satisfies \eqnref{eq:generalCondition} for small modulation strengths.
As \subfigref{fig:krein}{b} and \subfiglabel{d} illustrate, parametric resonances occur in pairs with one Floquet multiplier moving outward from the unit circle signaling amplification ($|\lambda(T)| > 1$) whereas its partner moves inward signaling damping ($|\lambda(T)| < 1$).

The collision of eigenvalues associated with partners of the same normal mode, eg. $e^{-i\Omega_i t} \to e^{+i \Omega_i t}$, occurs on the real axis at values of time $t \sim \inte \frac{\pi}{\Omega_i}$.
At even values of $\inte$, the eigenvalues meet at $+1$ (\subfigref{fig:krein}{a}), and the frequencies of the nascent amplified/damped modes are locked to the modulation frequency (see \appref{app:floquet}). 
At odd values of $\inte$, the collision happens at a value of $-1$, and the frequency of the modes is locked to double the modulation frequency.
These two cases typically lead to tangent (saddle-node) bifurcations and period-doubling bifurcations, respectively, when nonlinearity is added to the system~\cite{howard1987symplectic}.
If the eigenvalues associated with different normal modes meet, the collision of a pair of eigenvalues ($e^{-i\Omega_i t} \to e^{+i \Omega_j t}$ for example) is accompanied by a collision of the complex-conjugate eigenvalues, i.e., $e^{ +i\Omega_i t} \to e^{-i \Omega_j t}$ (as shown schematically in \subfigref{fig:krein}{c}).
We note that for a single oscillator with natural frequency $\Omega$, the resonance condition from Hill's equation of $T = \inte \pi/\Omega$ is recovered.

The generic resonance conditions described above hold in the limit of weak modulation, $\delta \to 0$.
As the modulation strength increases, resonances will generically be present for a range of modulation time-periods in the vicinity of $\inte \frac{2\pi}{\Omega_i+\Omega_j}$, opening regions of parametric resonance that grow with $\delta$~\cite{YakubovichStarzhinskii} (see also \figref{fig:twores}).
The size and shape of the regions away from $\delta=0$, and the strength of the resonances (i.e., the distance of the non-unitary eigenvalue of $U(T)$ from the unit circle), will depend on the specific functional form of the time-dependent modulation.
The ``instability tongues'' that appear in the vicinity of modulation periods $T= \frac{\inte\pi}{\Omega}$ for the Mathieu oscillator~\cite{kovacic2018mathieu} are a well-known example of these resonance regions.

Since parametric resonance corresponds to pseudo-Hermiticity breaking in $U(T)$, many techniques and phenomena in the theory of pseudo-Hermitian matrices carry over to the mechanical system under study. 
For example, stable phases (regions in a parameter space where all eigenvalues lie on the unit circle) of the Floquet matrix can be characterized topologically by the ordering of eigenvalues of positive and negative signature along the unit circle (compare \subfigref{fig:krein}{a} to \subfigref{fig:krein}{c}). 
It is impossible to traverse from one stable phase to a stable phase with a different ordering of eigenvalues without encountering an unstable degeneracy~\cite{melkani2023breaking}. 
Such topological phases may be harnessed for topologically protected behavior and the preparation of novel dynamical phases with no static analogs~\cite{wu2020phases, yu2021dynamical}.
Additionally, the boundaries, in parameter space, separating a stable phase from an amplified phase form lines or surfaces of exceptional points where eigenvectors of the Floquet matrix coalesce~\cite{melkani2023breaking}. 
These symmetry-protected exceptional points may be useful in the design of mechanical sensors~\cite{zhang2019noiseTheory} or in realizing topologically protected transport~\cite{xu2023heatTransport} and mode switching~\cite{arkhipov2023dynamically}.

\section{Effect of external symmetries}\label{sec:sym}

In the previous section, we linked the existence of parametric resonances to the formation of degeneracies in the Floquet matrix in the limit of vanishing parameter modulation.
This led to the resonance condition of \eqnref{eq:generalCondition} which is necessary, but not sufficient, for resonance to occur.
While a generic stiffness modulation with time period $T$ satisfying \eqnref{eq:generalCondition} is expected to lead to a resonance, the functional form of the time-modulation could be fine-tuned for a particular system to avoid resonances with any of its modes.
We expect these cases to occupy isolated points in the space of possible modulation functions, whose existence would be revealed by numerical evaluation of the Floquet exponents or by higher-order perturbation theory.
However, if we can identify external symmetries that prevent resonances from occurring, then we can rule out resonances for \emph{any} modulation that satisfies these external symmetries, without relying on fine-tuning.
The effect of external spatial and space-time symmetries on the parametric resonances of time-modulated features is addressed in this section.
While we focus on a model system of a ring of parametrically modulated oscillators for concreteness, the observations we make apply generally to time-modulated systems that satisfy spatial and space-time symmetries.

Consider the ring of oscillators in \figref{fig:coupledSystem}.
They are confined to oscillate in the vertical direction through the action of grounding springs of stiffness $k_j(t) = k[1+\delta f(t + j\tau)]$ and connected to their nearest neighbours via springs of zero rest length and static stiffness $g$~\footnote{The choice of zero rest length ensures that the spring applies a vertical restoring force of magnitude $g \Delta x$ when its ends have a relative displacement of $\Delta x$. Alternatively, springs under uniform tension $ga$, where $a$ is the separation of the two oscillators, also give rise to vertical restoring forces of the same magnitude.}.
Here, $f(t)$ is an arbitary periodic function with period $T$ and unit amplitude, and $0 \leq \tau < T$ quantifies the phase-shift in the modulation as we move from one oscillator to the next.
Our chosen modulation sets up a stiffness wave which travels from the $j$th oscillator to the $(j-1)$th oscillator over time (a direction we shall refer to as leftwards).

In the static limit $\delta \rightarrow 0$, this ring of oscillators has discrete translational symmetry.
The normal mode frequencies of the unmodulated system are $\pm \Omega(\kappa_1), \dots, \pm \Omega(\kappa_n)$ where $\kappa_j$ is the $j$th Bloch wavevector set by the system size and the periodic boundary conditions.
According to the discussion in \secref{subsec:Floquet}, upon turning on the parametric modulation we generically expect parametric resonance whenever the time-period of modulation $T$ is tuned to $T = \inte \frac{2\pi}{\Omega(\kappa_i)+\Omega(\kappa_j)}$ for some pair of normal mode frequencies.
We will consider two different kinds of symmetries and how they influence these conditions for parametric resonance.

\subsection{Discrete spatial symmetry in a modulated system ($\tau = 0$)}\label{subsec:sym1}

When the system is modulated at $\tau=0$, each oscillator undergoes the same modulation and the system retains translational symmetry at every point in time.
We can still apply Bloch's theorem to block-diagonalize the time-dependent Hamiltonian. 
The Floquet multipliers associated with the blocks corresponding to different Bloch wave-vectors decouple from each other.
As a result, parametric resonance only occurs at time periods equal to $T = \inte \frac{\pi}{\Omega(\kappa_i)}$, but not at $T = \inte \frac{2\pi}{\Omega(\kappa_i)+\Omega(\kappa_j)}$ with $i\neq j$.
At finite modulation amplitudes, the resonances open up a range of wave-vectors in the vicinity of the resonant $\kappa_i$ for which the corresponding Bloch waves become unstable~\cite{galiffi2019broadband, trainiti2019filtering}.

In fact, this decoupling due to symmetry is a general phenomenon~\cite{ge2014symmetry, melkani2023breaking}. 
Let $g_i$ be the elements of the group representation corresponding to a symmetry present in the system. 
Then $H(t)$ commutes with $g_i$ at all times $t$ and they share a common eigenbasis.
Now, for typical spatial symmetries (viz. rotation, translation, inversion, etc.) the elements $g_i$ take the form $\begin{pmatrix}
A & 0\\
0 & A
\end{pmatrix}$ where $A$ is some invertible square matrix (i.e., the matrices $g_i$ act on the positions and momenta in an equivalent manner). 
Any matrix of such form embodies a canonical transformation and hence commutes with the intertwining operator $G$.
Thus, $H(t)$ and $G$ can be block-diagonalized in the eigenbasis of the elements of the group representation. 
The system is then reduced to a set of uncoupled block matrices, each of which inherit the symmetries discussed in \secref{subsec:Floquet}.

\subsection{Space-time symmetry in a modulated system ($\tau = T/n$)}\label{subsec:sym2}

When the modulation at each oscillator is shifted in time by $\tau = \frac T n$ with respect to the oscillator on its left, the system enjoys space-time symmetry~\cite{xu2018spaceTimeGroup, gao2021spaceTimeTunneling, peng2022spaceTimeCrystal}. 
That is, the system is invariant after a translation by one position in the leftwards direction followed by a time evolution of $\tau$ time units.
Explicitly, the Hamiltonian satisfies
\begin{equation}\label{eq:hamiltonianSpaceTime}
H(t) = S H(t+T/n) S^{-1},
\end{equation}
where $S$ is the matrix operator which cyclically shifts each oscillator's coordinates by one position in the direction of the stiffness wave, i.e., to the left (see \appref{app:spacetime}).

For such symmetries, there is no way to block-diagonalize the time-dependent Hamiltonian. Nevertheless, we shall see that the system is still protected from parametric resonance at certain frequencies.
\eqnref{eq:hamiltonianSpaceTime} implies that the time-propagation matrix satisfies (see \appref{app:spacetime})
\begin{equation}\label{eq:unitarySpaceTime}
U(t+ T/n) = S^{-1} U(t) S U(T/n),
\end{equation}
which can be considered a generalization of Floquet's theorem (Floquet's theorem is recovered when $S = \mathbb I_{2n}$).
In particular, we have (using the fact that $S^n = \mathbb{I}_{2n}$) 
\begin{equation} \label{eq:sutn}
U(T)=S^{-n}[S U(T/n)]^n = [S U(T/n)]^n.
\end{equation}

The matrix $\spt:= S U(T/n)$, which we term the space-time Floquet matrix, enjoys the same internal symmetries discussed in \secref{subsec:Floquet}, i.e., it is real and satisfies $\spt^{-1} = G^{-1}\spt^\dagger G$.
(The latter equation relies on the fact that $S$ commutes with $G$ by the arguments in \secref{subsec:sym1}.)
As \eqnref{eq:sutn} shows, the Floquet matrix can be recovered from the space-time Floquet matrix.
We shall see below that the space-time Floquet matrix $\spt$ provides more information about the system dynamics, such as protected degeneracies and one-way amplification conditions, than the Floquet matrix $U(T)$ does.

\subsubsection{Protected degeneracies and avoided resonances} \label{sec:avoided}

It is possible for the Floquet matrix to have unstable degeneracies (where eigenvalues of opposite Krein signature collide) that do not arise in the space-time Floquet matrix $\spt$. 
Crucially, such a degeneracy in the Floquet matrix cannot develop into a parametric resonance (where Floquet multipliers move off the unit circle), as it would imply an instability in $\spt = [U(T)]^{1/n}$ without a corresponding degeneracy. 
To see this explicitly, consider the eigenvalues $\mu(T/n)$ of the space-time Floquet matrix in the limit $\delta \rightarrow 0$ when $\spt = SU(T/n) \sim Se^{-iH T/n}$. 
These eigenvalues are $e^{i\kappa_j}e^{\mp i \Omega(\kappa_j) T/n}$ and parametric resonance occurs when
\begin{equation}\label{eq:spacetimeCondition}
e^{i\kappa_j}e^{+ i \Omega(\kappa_j) T/n} = e^{i\kappa_k}e^{-i \Omega(\kappa_k) T/n}
\end{equation}
for some Bloch wave-vectors indexed by $j$ and $k$. 
These conditions are more restrictive than the general condition, \eqnref{eq:generalCondition}, which only considered internal symmetries.

For instance, consider the typical method of parametrically amplifying a mode indexed by $\kappa_i$ by modulating the system at $2\Omega(\kappa_i)$, i.e. modulation time period set to $T = \pi/\Omega(\kappa_i)$ which satisfies \eqnref{eq:generalCondition} with $r=1$.
As we saw in \secref{subsec:Floquet}, the underlying mechanism is the collision of eigenvalues of $U(T)$ associated with the two partners of the $i$th normal mode of the unmodulated system. 
However, for the space-time symmetric modulation, the collision of eigenvalues of $\spt$ is dictated by \eqnref{eq:spacetimeCondition}, which singles out $$T = \inte\frac{\pi n}{\Omega(\kappa_i)}$$ corresponding to modulation frequencies tuned to $\frac{2\Omega(\kappa_i)}{n}, \frac{2\Omega(\kappa_i)}{2n}, \dots$.
When the system size is large ($n \gg 1$), any possible instabilities due to the collision of two partners of the same normal mode will be protected from amplification.
This is because the first possible resonance occurs at modulation frequency tuned to $\frac{2\Omega(\kappa_j)}{n}$, and amplification at a such higher order resonance is generically much weaker~\cite{turner1998parametric}.

\begin{figure}[tb]
\includegraphics{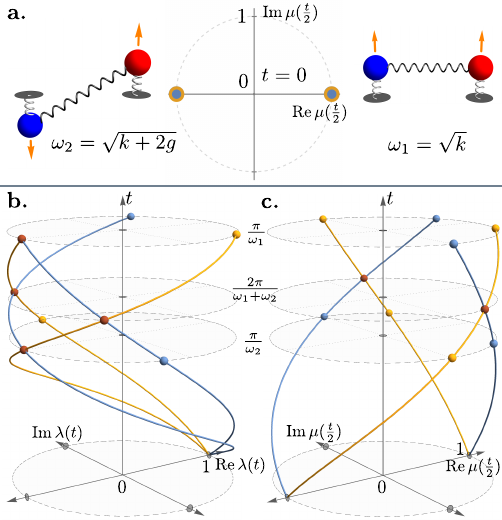}
\caption[]{\label{fig:spaceTime}
  Dimer ($n=2$) with space-time symmetry. 
  \subfiglabel{a.} Snapshots showing displacement and velocity of each mass in the anti-bonding (left) and bonding (right) modes. 
  Each mode contributes a pair of eigenvalues with opposite Krein signature to the spectrum of $SU(t/2)$ (center), shown as a blue disc and a yellow circle at $\mu=+1$ (bonding mode) and at $\mu = -1$ (anti-bonding mode) when $t=0$.
  We consider the static limit $\delta\rightarrow 0$ to find the conditions of parametric resonance.
  \subfiglabel{b.}   As time increases, the eigenvalues $\lambda(t)$ of $U(t) = e^{-iH t}$ of positive (negative) signature, shown in blue (yellow), move clockwise (counter-clockwise) along the unit circle starting from $\lambda(t=0)=+1$.
  Unstable degeneracies (red points) form at times equal to $\frac{\pi}{\omega_2}$, $\frac{2\pi}{\omega_1+\omega_2}$, and $\frac{\pi}{\omega_1}$ (and their integer multiples). 
\subfiglabel{c.} For a space-time symmetric modulation, the correct conditions for resonance are given by the matrix $SU(t/2)= S e^{-iH t/2}$ whose eigenvalues are $\mu(t/2)$. 
As the spectrum of $SU(t/2)$ evolves with time, unstable degeneracies occur only at $t = \frac{2\pi}{\omega_1+\omega_2}$.  At $t = \frac{\pi}{\omega_2}$ and $t = \frac{\pi}{\omega_1}$, even though the eigenvalues $\lambda(t)=[\mu(t/2)]^2$ are degenerate (at $-1$), the eigenvalues $\mu(t/2)$ are not (the corresponding values are $+i$ and $-i$).
}
\end{figure}

As an example, we analyze the case of $n=2$ oscillators modulated in a space-time symmetric manner (\figref{fig:spaceTime}), which corresponds to a modulation phase difference of $\pi$. 
Its two normal modes are the bonding mode where the two masses move in tandem ($\omega_1=\sqrt{k}$) and the anti-bonding mode where they move in opposite directions to each other ($\omega_2=\sqrt{k+2g}$) as illustrated in \subfigref{fig:spaceTime}{a}.
In the $\delta \to 0$ limit, the spectrum of $U(t)$ exhibits unstable degeneracies at $t=\inte \frac{\pi}{\omega_1}$, $t=\inte \frac{\pi}{\omega_2}$, and $t=\inte\frac{2\pi }{\omega_1 + \omega_2}$ (red dots in \subfigref{fig:spaceTime}{b} show the degeneracies when $r=1$), suggesting that the anti-bonding mode (for example) should resonate at modulation time-periods of $T= \inte \frac{\pi}{\omega_2}$.
However, the degeneracies in $SU(t/2)$ (\subfigref{fig:spaceTime}{c}) occur at $t= 2\inte\frac{\pi}{\omega_1}$, $t=2 \inte \frac{\pi}{\omega_2}$, and $t=(2\inte-1)\frac{2\pi }{\omega_1 + \omega_2}$ and thus the correct resonance condition for the anti-bonding mode is modulation at $T= \inte \frac{2 \pi}{\omega_2}$. 
Similarly, resonance due to the bonding mode coupling with the anti-bonding mode occurs at modulation time-periods $T= (2\inte-1)\frac{2\pi}{\omega_1 + \omega_2}$ in contrast to the prediction of $T=\inte\frac{2\pi }{\omega_1 + \omega_2}$ from the Floquet matrix degeneracies.
When the system is modulated at frequencies where parametric resonance is forbidden by the space-time symmetry, the Floquet matrix exhibits diabolic point degeneracies.

\begin{figure}
  \centering
  \includegraphics{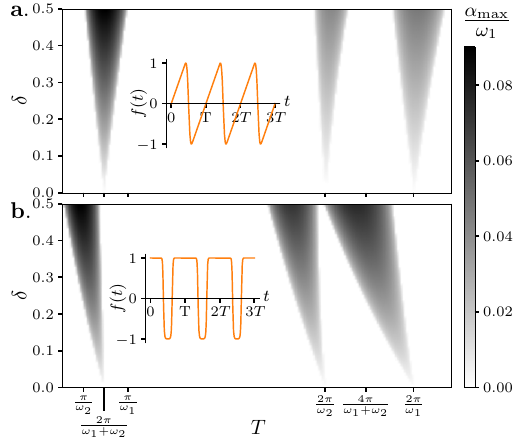}
  \caption{Numerically computed resonance regions for the space-time symmetric $n=2$ system with $k=1$ and $g=0.2$.
    The spring stiffnesses are modulated according to $k_j(t) = k\left[1+\delta f\left(t+j\frac{T}{2}\right)\right]$ with $j \in \{1,2\}$.
    Intensity plots of the largest gain factor in the Floquet spectrum, $\alpha_\text{max} = \text{max}(\alpha_i)$, are shown against modulation strength $\delta$ and modulation time period $T$ for two different time-modulation functions $f(t)$.
    Inset to each intensity plot shows the corresponding $f(t)$.
    Time periods corresponding to $\inte\frac{2\pi}{\omega_i+\omega_j}$ with $\inte=1$ and $\inte=2$ are indicated on the horizontal axis.
    \textbf{a.} Sawtooth-like stiffness modulation function generated via $f(t) = C \sin^{-1}\left[\tanh\left(10 \cos\left(\frac{\pi t}{T}\right)\right) \sin\left( \frac{\pi t}{T}\right)\right]$ with the constant $C$ chosen so that the wave amplitude is one.
    \textbf{b.} Rectangular wave-like modulation corresponding to $f(t) = \tanh\left[10\cos\left( \frac{2\pi t}{T}\right)+7\right]$. 
  }
    \label{fig:twores}
\end{figure}

To test these predictions, we numerically computed the Floquet multipliers of the $n=2$ system for two specific realizations of the time-modulation function, at different time periods and strengths (\figref{fig:twores}).
The existence of parametric resonance was established by measuring the gain factor corresponding to the $i$th Floquet multiplier, $\alpha_i \equiv \log \lambda_i/T$, which sets the long-time growth in amplitude of the corresponding Floquet eigenmode (see \appref{app:floquet}).
We observe regions of parametric resonance (nonzero gain factors) that originate from the predicted time-periods $T= \inte \frac{2 \pi}{\omega_1}$, $T= \inte \frac{2 \pi}{\omega_2}$ and $T = (2\inte-1)\frac{2\pi}{\omega_1 + \omega_2}$ with $r=1$.
As predicted by the space-time Floquet analysis, no resonances emerge from the less restrictive conditions, $t=\frac{\pi}{\omega_1}$, $t= \frac{\pi}{\omega_2}$, and $t=\frac{4\pi }{\omega_1 + \omega_2}$, which satisfy \eqnref{eq:generalCondition} but not \eqnref{eq:spacetimeCondition}.
A comparison of the sawtooth-like modulation function (\subfigref{fig:twores}{a}) to the rectangular wave modulation (\subfigref{fig:twores}{b}) confirms that the form of the modulation affects the strength of the parametric amplification and the shape of the resonance regions, but does not affect the resonance frequencies in the $\delta \to 0$ limit.

\subsubsection{Amplification of one-way travelling modes} \label{sec:oneway}

\begin{figure*}
\centering
\includegraphics{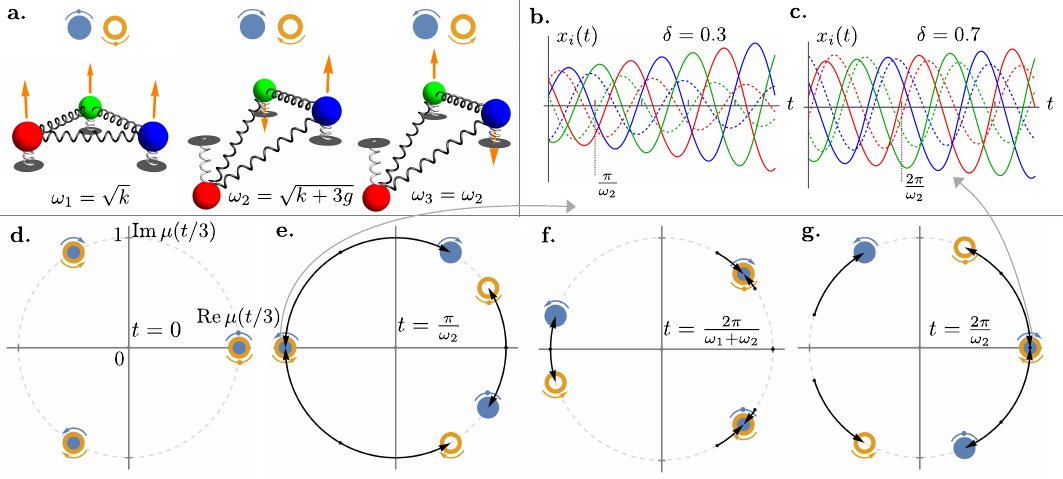}
\caption{\label{fig:oneWay}
  Trimer ($n=3$) with one-way propagating modes. The space-time symmetric modulation corresponds to a travelling wave of stiffness for the grounding springs with the maximum stiffness cycling in order of colors $R \to G\to B$ (clockwise).
  \subfiglabel{a.} Snapshots showing representative displacements and velocities of the masses for the three normal modes in the $\delta \to 0$ limit: a mode with all three masses oscillating in phase (left), and two degenerate travelling-wave modes with $\kappa = \pm 2\pi/3$---one travelling clockwise in the same direction as the stiffness wave (center), and the other travelling counter-clockwise (right).
  Symbols above the modes denote corresponding eigenvalues in panels \subfiglabel{d--g}, colored by Krein signature as in \figref{fig:krein}.
  \subfiglabel{b.} Numerically computed time traces of the oscillator positions when initialized to the amplified (continuous curve) and damped (dotted curve) Floquet eigenvectors (we used $k=10$, $g = 3$, $\delta = 0.3$, $T = \frac{\pi}{\omega_2}$, and $f(t)=\cos(2\pi t/T)$). 
  At this resonance, only the counterclockwise propagating mode resonates (peak follows $R \to B \to G$ with time). 
  \subfiglabel{c.} Same as \subfiglabel{b}, (but at $T = \frac{2\pi}{\omega_2}$ and $\delta=0.7$) for which only the clockwise mode experiences resonance.
  \subfiglabel{d.--g.} Eigenvalues $\mu(t/3)$ of $SU(t/3)= S e^{-iH t/3}$ in the static limit $\delta \rightarrow 0$ at values of time (labels) for which pairs of eigenvalues with opposite Krein signature collide.
}
\end{figure*}

The space-time Floquet matrix can also uncover the conditions for parametric resonance to occur selectively for modes that travel in one direction along the ring.
For a fixed Bloch wave-vector $0<\kappa_j<\pi$, the modes associated with $\Omega(\kappa_j)$ and $-\Omega(-\kappa_j)$ travel in the direction of the stiffness wave while the modes $-\Omega(\kappa_j)$ and $\Omega(-\kappa_j)$ travel in the opposite direction. 
(Note that $\Omega(\kappa_j)=\Omega(-\kappa_j)$ due to the $\mathcal{T}$-symmetry of the Hermitian system at $\delta \rightarrow 0$~\cite{susstrunk2016phonons}.)
When the pair $\Omega(\kappa_j)$ and $-\Omega(-\kappa_j)$ are coupled by the modulation, it leads to a pair of amplified and damped modes travelling in the leftwards direction whereas the rightwards travelling modes remain unamplified. 
Via \eqnref{eq:spacetimeCondition}, this coupling happens when the corresponding eigenvalues of $\spt$ collide, which occurs at modulation time-periods equal to
\begin{equation}\label{eq:oneWaycondition1}
T = n\frac{\pi (\inte-1) + \kappa_j}{\Omega(\kappa_j)}.
\end{equation}
By contrast, the right-wards moving modes (opposite to direction of stiffness wave) amplify at
\begin{equation}\label{eq:oneWaycondition2}
T = n\frac{\pi \inte - \kappa_j}{\Omega(\kappa_j)}.
\end{equation}

Remarkably, these two time-periods are different (except for the modes at wavevector $\kappa_j=\pi/2$).
Thus, equations~\eqref{eq:oneWaycondition1} and \eqref{eq:oneWaycondition2} provide simple criteria, using the symmetries of a system alone, to determine the conditions for the amplification of one-way travelling modes. 
Furthermore, one can control which direction of signal propagation is amplified just by tuning the modulation frequency in a space-time modulated system.

As a concrete example, consider the system of three oscillators whose normal modes in the absence of modulation are illustrated in \subfigref{fig:oneWay}{a}.
The space-time symmetric modulation generates a travelling wave in the clockwise sense when the masses are viewed from above.
The ground state mode, in which all masses oscillate in phase, resonates at a modulation frequency of $2\omega_1/3$ according to the analysis of \secref{sec:avoided}; the resonance at $2\omega_1$ is avoided even though it is the strongest resonance of an isolated individual oscillator.
The remaining two normal modes are degenerate and correspond to travelling waves in the clockwise (mode 2) and counterclockwise (mode 3) sense.
The space-time symmetric modulation breaks the degeneracy between these modes, as can be seen by following the eigenvalues of $SU(t/3)$ (\subfigref{fig:oneWay}{d--g}).
At $t=0$, each mode contributes one eigenvalue at $e^{2\pi i /3}$ and one  of opposite Krein signature at $e^{-2\pi i/3}$ (\subfigref{fig:oneWay}{d}).
As time advances, eigenvalues with different signatures travel in opposite clock senses, and the first pair of eigenvalues to collide are those from mode 3 (the mode propagating in the opposite direction as the stiffness wave) at $t=\pi/\omega_2$ (\subfigref{fig:oneWay}{e}).
This value is obtained from \eqnref{eq:oneWaycondition2} with $r = 1$ and $\kappa = -2\pi/3$.
The eigenvalues corresponding to the clockwise-propagating mode 2 collide at $t=2\pi/\omega_2$, obtained from \eqnref{eq:oneWaycondition1} with $r = 1$ and $\kappa = 2\pi/3$ (\subfigref{fig:oneWay}{g}).

These degeneracies in the spectrum of $SU(t/3)$ dictate the parametric resonances experienced by the travelling-wave modes when the system is periodically modulated.
When the modulation time-period is $T = \pi/\omega_2$ (modulation frequency twice the natural frequency of the travelling modes), only mode 3 experiences parametric resonance, contributing an amplified and a damped Floquet eigenvector.
The evolution of the oscillators when initialized to the Floquet eigenvectors are numerically computed and shown in \subfigref{fig:oneWay}{b}.
Since the eigenvalues collide at $-1$, the oscillatory component of these modes has a time period of twice the modulation period, which corresponds to the period of the original normal mode.
Mode 2, however, resonates at a different modulation period of $T = 2\pi / \omega_2$ (modulation frequency same as the mode frequency), with the eigenvalues colliding at $+1$, such that the period of the resulting oscillations equals the original mode periods (\subfigref{fig:oneWay}{c}).
The strength of amplification of this mode, quantified by the rate of change of the amplitude, is much weaker than that of the counterclockwise-propagating mode in \subfigref{fig:oneWay}{b}, since it is a higher-order resonance.

To summarize, when the system is modulated at twice the natural frequency of the degenerate travelling modes, only the mode travelling in opposite direction to the stiffness wave (mode 3) resonates.
At modulation frequency tuned to equal the natural frequency, the other mode in the degenerate pair (the previous mode's chiral partner) resonates.
These conditions are predicted by \eqnref{eq:oneWaycondition1} and \eqnref{eq:oneWaycondition2}.

The identification of one-way amplification in the ring of oscillators by the space-time Floquet matrix illustrates how it can predict the form of the resonant modes even when the normal modes are degenerate.
We note that the eigenvectors of $\spt$ are also eigenvectors of the Floquet matrix.
The converse, however, is true only when the Floquet multipliers are all non-degenerate. 
Consider again the three oscillator system in \figref{fig:oneWay}.
Since it has degenerate normal mode frequencies, the choice of the orthogonal normal mode basis vectors, and thus of the Floquet vectors in the static limit, is arbitrary.
Even though the degeneracy structure of the Floquet matrix correctly predicts some parametric resonances, such as $T=\frac{2\pi}{\omega_2}$, it does not identify which mode in the degenerate normal mode subspace will experience amplification/decay.
The eigenvectors of the space-time Floquet matrix, however, are non-degenerate and single out the two travelling modes as the physically relevant basis vectors for describing parametric amplification in the time-modulated system.

The breaking of chiral symmetry in the amplified modes is an example of reciprocity breaking in the system: wave propagation at the chiral resonances is highly asymmetric between the clockwise and counter-clockwise directions.
This non-reciprocity  would manifest as an asymmetry in the response of the other two masses when one mass is externally driven at the oscillatory frequency of the amplified mode.
Non-reciprocity of vibrational waves away from parametric resonances has been proposed and observed in three-port acoustic circulators~\cite{fleury2014nonreciprocity, goldsberry2020nonreciprocity}.
In these prior works, the non-reciprocal response was due to a splitting in the oscillation frequencies of the clockwise and counter-clockwise modes due to a traveling-wave modulation.
In our system, not only do the oscillation frequencies (related to the phase of the Floquet multipliers, see \appref{app:floquet}) move away from each other, but the frequency of one of the traveling-wave modes becomes complex (signifying amplification) while the other remains real.
This represents a qualitatively different form of reciprocity breaking tied to one-way amplification, which could be used to engineer mechanical amplifiers with favorable noise characteristics~\cite{rugar1991squeezing, metelmann2015nonreciprocal}.

\subsection{Generalization to arbitrary space-time symmetric modulation}\label{subsec:sym3}

So far, we considered space-time symmetric modulation in the ring of coupled parametric oscillators with the specific phase shift $\tau=T/n$.
We now briefly address general modulation patterns obtained by considering other values of $\tau$.

If the phase shift is incommensurate with the periodic boundary conditions, i.e., $n\tau \operatorname{mod} T \neq 0$, then the system hosts defects in the space-time modulation pattern and will be outside the scope of this work. 
Such a system may be analyzed using the theory of non-Hermitian Floquet defects and may exhibit the skin effect~\cite{lin2023topological, gao2022skinOscilllators}.

To retain periodic boundary conditions, we must have $n\tau \operatorname{mod} T = 0$, or equivalently $\tau = \frac{p}{q}T$ where $p$ and $q$ are positive integers with no common factors and $q$ divides $n$.
This constraint allows for the following two cases.
First, if $q=n$, then our present analysis can still be applied to such a system, eg. $n=7$ oscillators with a phase shift of $\tau=\frac{3}{7}T$.
Explicitly, the Floquet matrix will now be factorized as $U(pT) = U(T)^p = [SU(\tau)]^n = [SU\left (\frac{p}{n}T\right )]^n$.

Second, we may have $q<n$, eg. $n=24$ oscillators with a phase shift of $\tau=\frac{5}{8}T$.
Such a system exhibits both the static translational symmetry at all times as well as space-time symmetry, i.e., it is a Floquet-Bloch lattice of $n/q$ supercells each with $q$ oscillators. 
Now the basis of vectors which are invariant on a translation by one oscillator (viz. the eigenvectors of $S$) not only diagonalize $S$, they also block-diagonalize the time-dependent Hamiltonian (since they are, in particular, also invariant on a translation by $n/q$ oscillators). 
In such a basis, the matrices $H(t)$, $U(t)$, and $S$  in \eqnref{eq:hamiltonianSpaceTime} and \eqnref{eq:unitarySpaceTime} are all block-diagonal and thus the space-time symmetry procedure can be applied to each block individually. 
Thus, while our framework is still applicable, the interplay of space-time symmetry with a Brillouin zone generates additional structure, such as winding numbers and bandgaps~\cite{gao2021spaceTimeTunneling}, which lie outside the scope of the current work and will be the subject of a future paper.

\section{Discussion}\label{sec:discussion}

In this work, we identified the non-Hermitian internal symmetries of the evolution operators governing the dynamics of a time-dependent linear classical mechanical system.
Using these internal symmetries, which arise from the real-valuedness and symplecticity of the equations of motion, we derived the parametric resonance conditions for arbitrary periodically modulated coupled oscillator systems.
These conditions are heralded by the collision of eigenvalues of opposite signature in the spectrum of the Floquet matrix---which can be exactly solved in the static limit.

We then examine systems with external symmetries, especially combined space-time symmetries.
By proposing a new framework for space-time symmetric systems, in terms of the space-time Floquet matrix, we find the conditions for these protected degeneracies (which lead to modes that are protected from parametric resonance and do not exhibit amplification or decay).
These conditions are formulated from the symmetries of the system alone without relying on the functional form of the modulation.

While protected degeneracies have been observed in the Floquet spectra of quantum as well as classical systems with time-modulated parameters (in the two-oscillator system~\cite{grossmann1991tunneling, breuer1988avoided, calvanese2019coupled, bello2019coherentBeating} and in space-time symmetric lattices~\cite{xu2018spaceTimeGroup}), their origins were not fully understood previously.
We show that protected degeneracies form when the space-time Floquet matrix, which captures the long-time behavior of a space-time symmetric system, is raised to an integer power to get the ordinary Floquet matrix.
If distinct eigenvalues of the former matrix become degenerate in the latter, they are protected from resonance.

Furthermore, our framework allows us to solve for the conditions for one-way amplification in a system with periodic boundary conditions, such as a ring of oscillators.
Remarkably, these conditions show that one can control which direction of signal propagation is amplified simply by tuning the modulation frequency in a space-time symmetric system, representing a strong breaking of reciprocity in the system dynamics.
Amplification of one-way modes has been seen in non-Hermitian Floquet systems before in both numerics~\cite{kruss2022oneway, adlakha2023demand} and experiments~\cite{wen2022unidirectional}. 
Our analysis, specifically through Equations~\eqref{eq:oneWaycondition1} and \eqref{eq:oneWaycondition2}, provides general conditions for the existence of one-way amplification in terms of the mode structure in the static limit.
In contrast to approaches such as in Ref.~\onlinecite{koutserimpas2018nonreciprocalGain}, we do not rely on any particular forms of the modulation function to generate directional amplification.
Indeed, all of our analysis depends only on the symmetries of the system and applies to arbitrary functional forms of the time modulation.
Our framework for space-time symmetry is applicable to all systems, Hermitian as well as non-Hermitian, with such symmetry~\cite{coulais2021broken}.

We expect the insights provided by our theoretical study to enable the engineering of amplification, unidirectional transport, as well as protected eigenvalue degeneracies in space-time modulated materials.
Since the onset of parametric resonance is triggered by an exceptional degeneracy~\cite{melkani2023breaking}, our results can also be used to engineer sensing~\cite{zhang2019noiseTheory} and mode-switching devices~\cite{arkhipov2023dynamically} based on exceptional-point physics.
The one-way amplified modes could also serve as the basis for limit cycles that do useful mechanical work in the presence of nonlinearities~\cite{brandenbourger2022limit}.
The formulation based on the space-time Floquet matrix shows that space-time symmetric systems may be used to realize non-trivial $n$-root analogs of systems with topologically protected states~\cite{arkinstall2017squareRoot,marques2021nRoot}.
Other promising directions for future work include a detailed analysis of the interplay of space-time symmetry with band structures in systems where the space-time modulation leads to a lattice of supercells, and the influence of open boundary conditions which can cause dramatic changes in non-Hermitian spectra~\cite{yao2018edge}.

\begin{acknowledgments}
  \acktext
\end{acknowledgments}

\appendix
\setcounter{figure}{0}
\renewcommand{\thefigure}{A\arabic{figure}}

\section{Generalizations to other mechanical systems}\label{app:generalizations}

\subsection{Hamiltonian approach to deriving symmetries}

A classical mechanical system is described by $n$ coordinates $x_i$ and $n$ canonically conjugate momenta $p_i$. The classical-mechanical Hamiltonian $\mathcal H(\bs x, \bs p)$ governs the dynamics via the equations
\begin{equation}
\frac{d x_i}{dt} = +\frac{\partial \mathcal H}{\partial p_i} \quad \textrm{and }
\frac{d p_i}{dt} = -\frac{\partial \mathcal H}{\partial x_i} 
\end{equation}
which can be written in matrix form as
\begin{equation}
\frac{d}{dt} \begin{pmatrix} \bs x \\ \bs p \end{pmatrix}
= \begin{pmatrix} \bigz & \id \\ -\id & \bigz \end{pmatrix}
\begin{pmatrix} \nabla_{\bs x} \mathcal H \\ \nabla_{\bs p} \mathcal H \end{pmatrix}.
\end{equation}

If the Hamiltonian is a quadratic function, the equations of motion would be linear. A general quadratic Hamiltonian is
\begin{equation}
\mathcal H = \sum_{i,j}\left ( \frac{a_{ij}}{2}x_i x_j + \frac{b_{ij}}{2} p_i p_j + c_{ij} x_i p_j \right)
\end{equation}
where $a_{ij} = a_{ji}$ and $b_{ij}=b_{ji}$. The coefficients above, which may in general be time-dependent, define the real matrices $A$, $B$, and $C$, where $A$ and $B$ are symmetric. For such a Hamiltonian we have
\begin{equation}
\begin{pmatrix} \nabla_{\bs x} \mathcal H \\ \nabla_{\bs p} \mathcal H \end{pmatrix}
=\begin{pmatrix} A & C\\ C^T & B \end{pmatrix}
\begin{pmatrix} \bs x \\ \bs p \end{pmatrix}.
\end{equation}

Our linear system is then
\begin{equation}\label{eq:generalHamiltonian}
\frac{d}{dt} \begin{pmatrix} \bs x \\ \bs p \end{pmatrix}
= \begin{pmatrix} C^T & B \\ -A & -C \end{pmatrix}
\begin{pmatrix} \bs x \\ \bs p \end{pmatrix}.
\end{equation}
All linear systems which can be written in this form enjoy the internal symmetries discussed in this work.
We show below how mechanical systems with velocity-dependent forces or dissipation can be accomodated, under some assumptions, to match the form above. 

\subsection{Effect of gyroscopic forces}

In the presence of (time-independent) gyroscopic forces, the equations of motion generalize to~\cite{susstrunk2016phonons} 
\begin{equation}
\ddot{\bs x} = -K(t) \bs x + \Gamma \dot{ \bs x}.
\end{equation}
Here, $\Gamma$ is a real and skew-symmetric matrix ($\Gamma^T = -\Gamma$) which accounts for the frictionless forces that break reciprocity, such as the Lorentz force and the Coriolis force. Such forces take the form $\vec F = \vec \Omega \times \dot{\vec x} = \sum_{jkl} \epsilon_{jkl} \Omega_j \dot{x}_k \hat{e}_l$ where $\sum_j\epsilon_{jkl}\Omega_j$ indeed reverses its sign upon the interchange
of $k$ and $l$~\cite{ashida2020nonhermitian}.

To reach the desired form we define the vector $\bs v = \dot{\bs x} - \frac{\Gamma}{2} \bs x$ such that
\begin{equation}
    \dot{\bs v} = \left (- K(t) + \frac{\Gamma^2}{4} \right ) \bs x + \frac{\Gamma}{2} \bs v.
\end{equation}

Thus,
\begin{equation}
\begin{pmatrix} \dot{\bs x} \\ \dot{\bs v} \end{pmatrix} = 
    \begin{pmatrix} \frac{\Gamma}{2} & I_{n} \\ -K(t) + \frac{\Gamma^2}{4}& \frac{\Gamma}{2} \end{pmatrix}
    \begin{pmatrix} \bs x \\ \bs v \end{pmatrix},
\end{equation}
which is of the same form as \eqnref{eq:generalHamiltonian}.

\subsection{Effect of dissipation due to friction}\label{app:dissipation}

In the presence of dissipation due to friction the Hamiltonian changes to
\begin{equation}
H_{\text{d}}(t)= -i\begin{pmatrix} 
\bigz & -\id \\ 
K(t) & \gamma 
\end{pmatrix}
\end{equation}
where $\gamma$ is a diagonal matrix with the $j^{\mathrm{th}}$ term in the diagonal being the dissipation constant corresponding to the $j^{\mathrm{th}}$ momenta. The trace-less matrix
\begin{equation}
    \Tilde H(t) = H_{\text{d}}(t) +\frac{i}{2}\begin{pmatrix} \gamma & \bigz \\ \bigz & \gamma \end{pmatrix} = -i\begin{pmatrix} -\frac{1}{2}\gamma & -\id \\ K(t) & \frac{1}{2}\gamma \end{pmatrix}
\end{equation}
has the same symmetries as the frictionless Hamiltonian $H(t)$ in \eqnref{eq:timeDependent} of the main text. That is, $\Tilde H(t)$ is also purely imaginary and it also satisfies the pseudo-Hermiticity condition, $G \Tilde H(t) G^{-1} = \Tilde H^\dagger(t)$, as in \eqnref{eq:symplectic} of the main text~\cite{yoshida2019ring}. Essentially, we converted $H_{\text{d}}(t)$ to a matrix $\Tilde H(t)$ with balanced gain and loss.

However, for such a transformation to be permissible in a time-dependent system, we have to assume that the dissipation is uniform, i.e., the dissipation coefficient is the same for all oscillators such that $\Tilde H(t) = H_{\text{d}}(t) +\frac{i}{2}\gamma \mathbb{I}_{2n}$ where $\gamma$ now denotes a real number. In this case, we can transform our coordinates via 
\begin{equation}
\begin{pmatrix}
\bs{\Tilde x}(t)\\
\bs{\Tilde p}(t)
\end{pmatrix} = 
e^{\frac{\gamma t}{2}}\begin{pmatrix}
\bs x(t)\\
\bs p(t)
\end{pmatrix}
\end{equation}
to get as a new equation of motion
\begin{equation}
i\frac{d}{dt}\begin{pmatrix}
\bs{\Tilde x}(t)\\
\bs{\Tilde p}(t)
\end{pmatrix} = \Tilde H(t)
\begin{pmatrix}
\bs{\Tilde x}(t)\\
\bs{\Tilde p}(t)
\end{pmatrix}.
\end{equation}

\section{Review of Floquet methods}\label{app:floquet}

\subsection{Long-time behavior of system}

The equation $U(t+ T) = U(t) U(T)$ can be derived as (with $\mathcal T$ being the time-ordering operator),
\begin{equation}\label{eq:floquetDerivation}
\begin{aligned}
U(t+T) &= \mathcal{T}\left [ \exp \left( -i\int_0^{t+T} dt' H(t') \right ) \right ]\\
&= \mathcal{T}\left [ \exp \left( -i\int_{T}^{t+T} dt' H(t') \right ) \right ] U(T)\\
&= \mathcal{T}\left [ \exp \left( -i\int_{0}^{t} dt' H(t') \right ) \right ] U(T)\\
&= U(t) U(T).
\end{aligned}
\end{equation}
This implies $U(mT) = U(T)^m$ where $m$ is any positive integer.

Given initial conditions $\{\bs x(0), \bs p(0)\}$, the coordinates of the system after $m$ time-periods is given by
\begin{equation}
\begin{pmatrix} \bs x(mT) \\ \bs p(mT) \end{pmatrix} = U(mT) \begin{pmatrix} \bs x(0) \\ \bs p(0) \end{pmatrix} = U(T)^m \begin{pmatrix} \bs x(0) \\ \bs p(0) \end{pmatrix}.
\end{equation}
Let $\ket{v_i}$ be the eigenvectors of $U(T)$, i.e. the Floquet eigenvectors, with eigenvalues $\lambda_i$, the Floquet multipliers. 
If the Floquet eigenvectors span all space, we can write the initial conditions as a superposition of these eigenvectors.
\begin{equation}
\begin{pmatrix} \bs x(0) \\ \bs p(0) \end{pmatrix} = \sum_i \alpha_i \ket{v_i},
\end{equation}
such that the equation above reduces to
\begin{equation}
\begin{pmatrix} \bs x(mT) \\ \bs p(mT) \end{pmatrix} = \sum_i \alpha_i \lambda_i^m \ket{v_i}.
\end{equation}
The Floquet multipliers then determine the long time behavior of the system.
A similar statement is true even in the case of an exceptional point when the Floquet eigenvectors do not span the whole space~\cite{YakubovichStarzhinskii}. 
For example, let $U(T)$ have a two-fold exceptional point degeneracy at $\lambda_j$. 
We can then define the generalized eigenvector $\ket{w_j}$ at $\lambda_j$ by
\begin{equation}
\begin{aligned}
U(T)\ket{v_j} &= \lambda_j \ket{v_j},\\
U(T)\ket{w_j} &= \lambda_j \ket{w_j} + \ket{v_j}.
\end{aligned}
\end{equation}
The vectors $\{\ket{w_j}, \ket{v_1},\dots, \ket{v_{n-1}}\}$ span all space and we can write our initial conditions as
\begin{equation}
\begin{pmatrix} \bs x(0) \\ \bs p(0) \end{pmatrix} = \sum_{i} \alpha_i \ket{v_i} + \beta_j \ket{w_j},
\end{equation}
such that
\begin{equation}
\begin{pmatrix} \bs x(mT) \\ \bs p(mT) \end{pmatrix} = \sum_i \alpha_i \lambda_i^m \ket{v_i} + \beta_j \left( \lambda_j^m \ket{w_j}+ m\lambda_j^{m-1} \ket{v_j} \right).
\end{equation}

\subsection{Frequency of Floquet modes}

When initial conditions of the oscillators are set to one of the Floquet eigenvectors, we call the ensuing motion a Floquet mode.
We saw above that, in the absence of exceptional points, the dynamics of the system can be decomposed into the Floquet modes much like how the dynamics of a static system can be decomposed into its normal modes.
The frequency of oscillations of a Floquet mode depends both on the modulation frequency and the Floquet multiplier for that eigenvector. 
For example, with initial conditions $\begin{pmatrix} \bs x(0) \\ \bs p(0) \end{pmatrix} = \ket{v_i}$, the system evolves as
\begin{equation}
\begin{pmatrix} \bs x(t) \\ \bs p(t) \end{pmatrix} = U(t) \ket{v_i} = U(mT + t_0) \ket{v_i} = U(t_0) \lambda_i^m \ket{v_i}.
\end{equation}
Here, $t=mT+t_0$ with $m$ a non-negative integer and $0 \leq t_0 < T$. 
Expressing, the Floquet multiplier as $\lambda_i= e^{(\alpha_i -i\omega_i)T}$ with $\alpha_i$ and $\omega_i$ real, we have
\begin{equation}
\begin{pmatrix} \bs x(t) \\ \bs p(t) \end{pmatrix} = U(t_0) e^{-i m \omega_i T} e^{m \alpha_i T}\ket{v_i}.
\end{equation}

We see that the coordinates of the oscillators return to the scaled value of their initial coordinates when $U(t_0)$ equals identity and $e^{-i m \omega_i T}$ equals one. 
The first time this happens is at time equal to the least common multiple of the modulation time-period $T$ and $\frac{2\pi}{\omega_i}$. 
This least common multiple is the time-period of the Floquet mode.

We consider two specific cases. 
When parametric amplification occurs due to a Krein collision of eigenvalues at $+1$, the Floquet multipliers for the nascent modes have $\omega_i=0$. 
The frequency of these modes are then locked to the modulation frequency. 
On the other hand, when the collision occurs at $-1$, the Floquet multipliers for the nascent modes have $\omega_i=\frac{\pi}{T}$, and their frequency is locked to twice the modulation frequency. 

\section{Derivation of generalized Floquet theorem in the presence of space-time symmetry}\label{app:spacetime}

The matrix $S$, which cyclically shifts each oscillator's coordinates by one position to the left, satisfies
\begin{equation}
S \begin{pmatrix}
x_1\\
x_2\\
\vdots\\
x_n\\
p_1\\
p_2\\
\vdots\\
p_n
\end{pmatrix} = 
\begin{pmatrix}
x_2\\
\vdots\\
x_{n}\\
x_1\\
p_2\\
\vdots\\
p_{n}\\
p_1
\end{pmatrix}.
\end{equation} 

For the time-propogation operator, we have (with $\mathcal T$ being the time-ordering operator),
\begin{align}\label{eq:spaceTimeDerivation}
U(t+ T/n) &= \mathcal{T}\left [ \exp \left( -i\int_0^{t+T/n} dt' H(t') \right ) \right ] \nonumber \\
&= \mathcal{T}\left [ \exp \left( -i\int_{T/n}^{t+T/n} dt' H(t') \right ) \right ] U(T/n)\nonumber \\
&= \mathcal{T}\left [ \exp \left( -i\int_{0}^{t} dt' S^{-1} H(t') S  \right ) \right ] U(T/n)\nonumber \\
&= S^{-1} U(t) S U(T/n).
\end{align}

The derivation for Floquet systems in \secref{app:floquet} is a special case of this derivation above.

\end{document}